\documentclass[a4paper,fleqn,usenatbib]{mnras}

\usepackage[T1]{fontenc}
\usepackage{ae,aecompl}

\usepackage{graphicx}	
\usepackage{amsmath}	
\usepackage{amssymb}	
\usepackage{relsize}

\newcommand{\msun}{\hbox{M$_{\odot}$}}
\newcommand{\kms}{\hbox{km s$^{-1}$}}

\title[Rejuvenation in AGN galaxies]{Rejuvenation triggers nuclear activity in nearby galaxies}

\author[I. Mart\'in-Navarro]{Ignacio Mart\'in-Navarro$^{1,2}$\thanks{E-mail: imartin@iac.es}, Francesco Shankar$^{3}$,  Mar Mezcua$^{4,5}$
\\
$^{1}$Instituto de Astrof\'isica de Canarias, V\'ia L\'actea s/n, E-38205 La Laguna, Tenerife, Spain\\
$^{2}$Departamento de Astrof\'isica, Universidad de La Laguna, E-38205 La Laguna, Tenerife, Spain\\
$^{3}$Department of Physics and Astronomy, University of Southampton, Highfield, SO17 1BJ, UK\\
$^{4}$Institute of Space Sciences (ICE, CSCIC), Campus UAB, Carrer de Can Magrans, 08193, Barcelona, Spain\\
$^{5}$Institut d'Estudis Espacials de Catalunya (IEEC), C/ Gran Capit\`{a}, 08034 Barcelona, Spain
}

\date{Accepted XXX. Received YYY; in original form ZZZ}

\pubyear{2017}

\begin{document}
\label{firstpage}
\pagerange{\pageref{firstpage}--\pageref{lastpage}}
\maketitle

\begin{abstract}
Feedback, in particular from active galactic nuclei (AGN), is believed to play a crucial role in the evolution of galaxies. In the local Universe, many galaxies with an AGN are indeed observed to reside in the so-called green valley, usually interpreted as a transition phase from a blue star-forming to a red quenched state. We use data from the Sloan Digital Sky Survey to show that such an interpretation requires substantial revision. Optically-selected nearby AGN galaxies follow exponentially declining star formation histories, as normal galaxies of similar stellar and dark matter halo mass, reaching in the recent past ($\sim$0.1 Gyr ago) star formation rate levels consistent with a quiescent population. However, we find that local AGN galaxies have experienced a sudden increase in their star formation rate, unfolding on timescales similar to those typical of AGN activity, suggesting that both star formation and AGN activity were triggered simultaneously. We find that this quenching followed by an enhancement in the star formation rate is common to AGN galaxies and more pronounced in early type galaxies. Our results demonstrate that local AGN galaxies are not just a simple transition type between star-forming and quiescent galaxies as previously postulated.
\end{abstract}



\section{Introduction} \label{sec:intro}

Galaxies, forming and evolving within their host dark matter haloes, are the end-product of a balance between gas cooling, star formation and feedback \citep[e.g.][]{Nelson19,Dav2020,Mitchell20}. However, understanding the details of the mechanisms driving the formation and evolution of galaxies is still an open and unsolved issue. In particular, it remains unclear how galaxies transition from a star-forming to a quenched state \citep[][]{Schawinski07,Schawinski14,Bluck20,Angthopo20}. One of the most popular processes invoked to cease or reduce star formation is the so-called energetic/momentum AGN feedback \citep[e.g.][]{Terrazas20}. A supermassive black hole accreting gas from the surroundings can in fact generate enough energy/momentum to heat up/expel the gas from the host galaxy via winds and/or jets. Observational evidence in the local Universe in support of this process was put forward in the past by preliminary results showing that local galaxies hosting AGN are preferentially in the so-called ``green'' valley, i.e. they indeed appear as galaxies transitioning from a blue star-forming sequence to a red and dead phase. In the present study we revisit this hypothesis by carefully and homogeneously analyzing a large Sloan Digital Sky Server \citep{SDSS10} (SDSS) sample of nearby active galaxies. 

\section{Sample and Analysis} \label{sec:sample}

The sample was selected as follows. First, we compiled publicly-available black hole masses derived from SDSS single-epoch optical spectroscopy of Type I AGN \citep{Greene07,Dong12,Reines15,Woo15,Chilingarian18,Liu19}. We then cross-matched this initial sample with a catalog of groups and clusters, also identified in SDSS \citep{Lim17}, with estimated dark matter halo masses. We limited our analysis to central galaxies. 

\begin{figure} 
    \begin{center}
    \includegraphics[width=8.5cm]{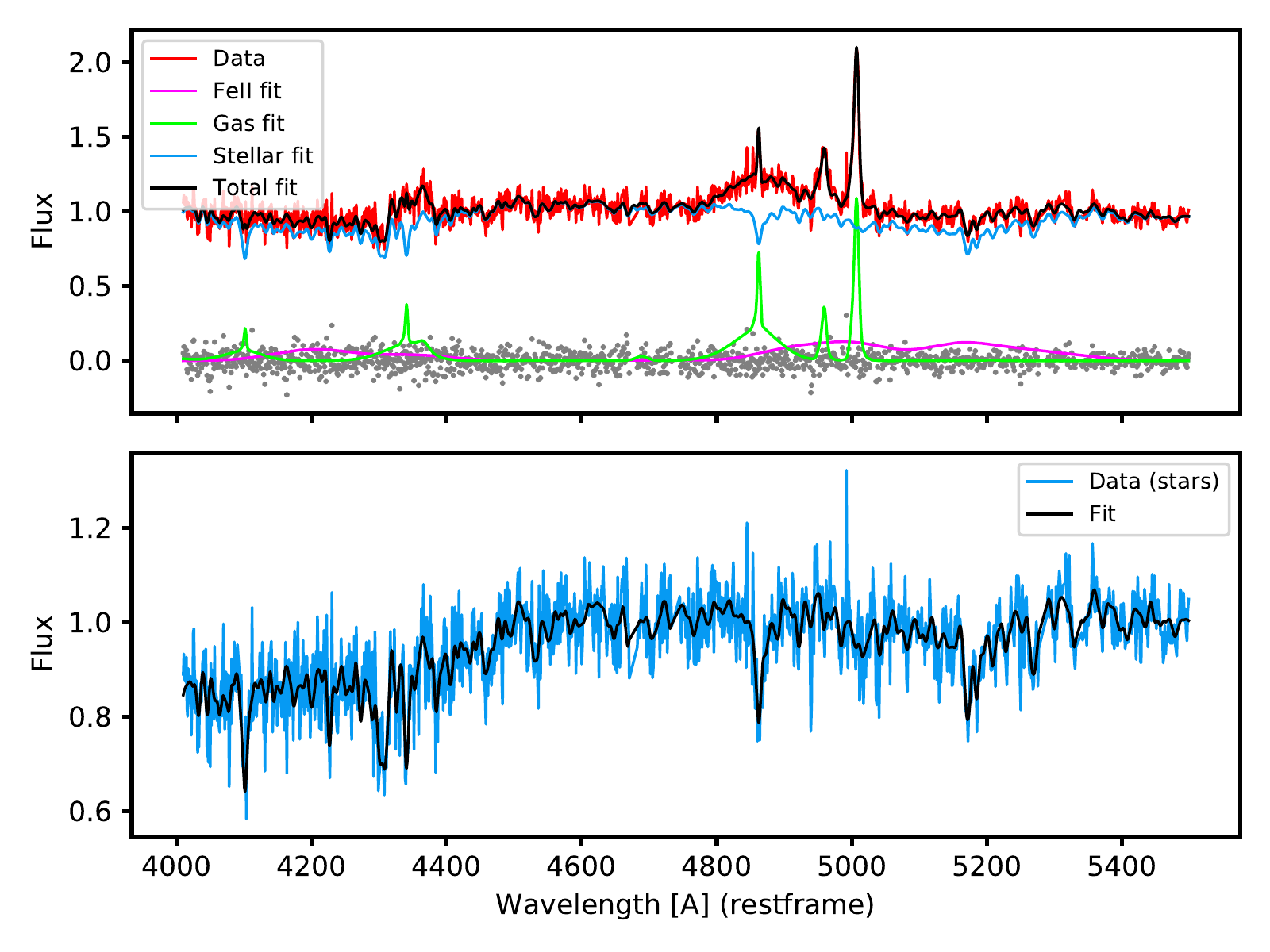}
    \end{center}
    \caption{{\bf Optical SDSS spectra and best-fitting model}. The red line in the top panel shows the observed SDSS data over the fitted wavelength range. We show the best-fitting FeII and gas emission templates in fuchsia and green, respectively, and in blue the stellar absorption spectra. Black line indicates the total best-fitting model while the gray dots show the fitting residuals. The bottom panel show a closer look to the stellar absorption spectra after corrected from the emission lines (in blue), with the best-fitting model shown in black.}
    \label{fig:0}
\end{figure}

Finally, we measured the stellar mass growth in our sample by fitting their optical spectra with a regularized linear combination \citep{ppxf} of single stellar population models \citep{Vazdekis15}. Our choice of these models is motivated by their use of empirical stellar spectra over the wide range of ages and metallicities expected for our sample. With this approach we recovered, in a non-parametric way, the star formation history of each galaxy in the sample. Given the best-fitting model, we also derived the expected mass-to-light ratios and thus the total stellar masses of our galaxies based on their k-corrected $r$-band photometry \citep{Blanton05,Padmanabhan08}. In addition to the stellar continuum, we also allowed pPXF to fit for emission lines in case they are needed to improve the best-fitting solution. Narrow lines were fit using two kinematically independent components to properly model the strongest emission lines, a third kinematical component was included to fit the broad Balmer emission. Emission from ionized iron lines \citep{feii} was also included as a separate kinematic component.

Our stellar population fitting focused on rest-frame wavelengths from $\lambda=4000$ \ \AA \ to $\lambda=5500$ \ \AA, a range where population synthesis models are more reliable and stellar population parameters can be more easily measured. Including wavelengths bluewards of $\lambda=4000$\,\AA \, could, in principle, improve our sensitivity to young stellar populations. However, in practice, the presence of strong emission lines in that region made our stellar population less robust. Moreover, beyond $\lambda\sim 5500$\,\AA, stellar population models become more sensitive to the atmospheres of cool stars and thus, prone to be affected by systematics related to both changes in the initial mass function \citep[e.g.][]{conroy} and non-canonical contributions from AGB stars \citep[e.g.][]{Maraston05}. We also masked out those wavelengths affected by strong telluric lines. An example of the SDSS spectrum and best-fitting model is shown in Fig.~\ref{fig:0}.

In summary, our final sample consists of 3,314 central, currently active galaxies, with known halo, black hole, and stellar masses, and for which we derived detailed, non-parametric star formation histories from their integrated optical spectra. While these selection criteria restricted the number of objects in our final galaxy and might lead to subtle biases due to e.g. orientation and dust obscuration\footnote{Although dust attenuation is in practice modelled during the stellar population fitting as a multiplicative component.}, they allowed us to assess the interplay between three fundamental ingredients in galaxy evolution: dark matter halos, black holes, and baryons (as probed by the stellar component). The main properties of our sample, covering a range in stellar masses from $\log$\,M$_\star=10.0$ to 10.8 $\log$\,\msun \ (10$th$ and 90$th$ percentiles, respectively), are listed in Table~\ref{tab:1}.

For reference, we also selected a control sample of central galaxies with same stellar mass distribution as our main sample but with no optical AGN-like line-ratios (i.e. excluding also Type II AGNs). This control sample allowed us to investigate whether the formation history of galaxies currently exhibiting nuclear activity (as revealed by their broad line emission) differs from the general population of galaxies with the same stellar mass. 

\section{Results} \label{sec:resu}

\begin{figure} 
    \begin{center}
    \includegraphics[width=8.5cm]{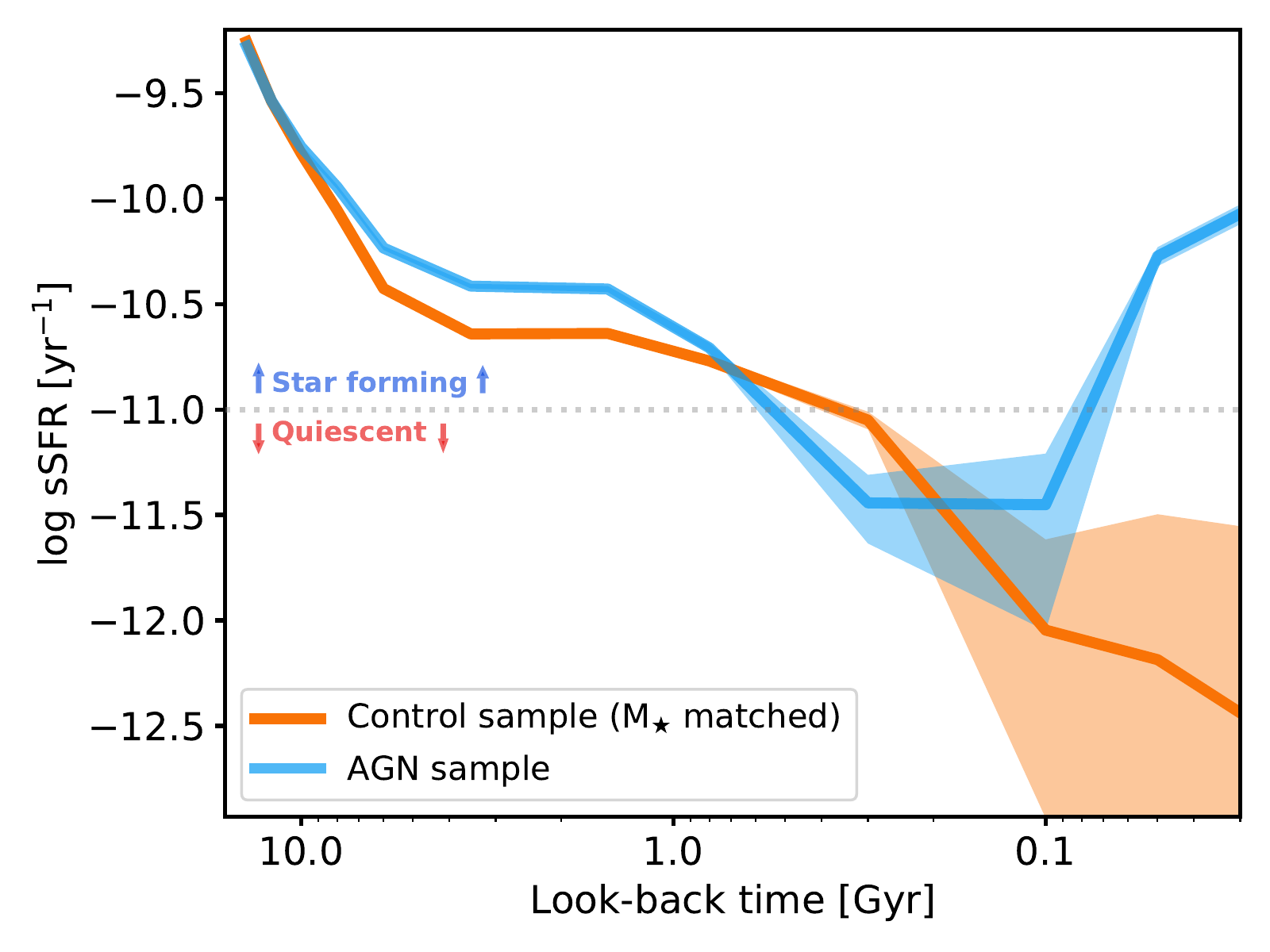}
    \end{center}
    \caption{{\bf Rejuvenation of active galaxies.} Evolution of the sSFR for our sample of active (blue) and control (orange) samples. The sSFR of both AGN and control samples has continuously decreased for more than 10 Gyr, reaching values consistent with a quiescent population ($\log$ sSFR $<-11$yr$^{-1}$). However, our AGN sample has experienced
    a recent (last $\sim 0.1$~Gyr) increase in the sSFR. Shaded areas indicate the 1$\sigma$ confidence interval.}
    \label{fig:1}
\end{figure}

\begin{figure*} 
    \begin{center}
    \includegraphics[width=17cm]{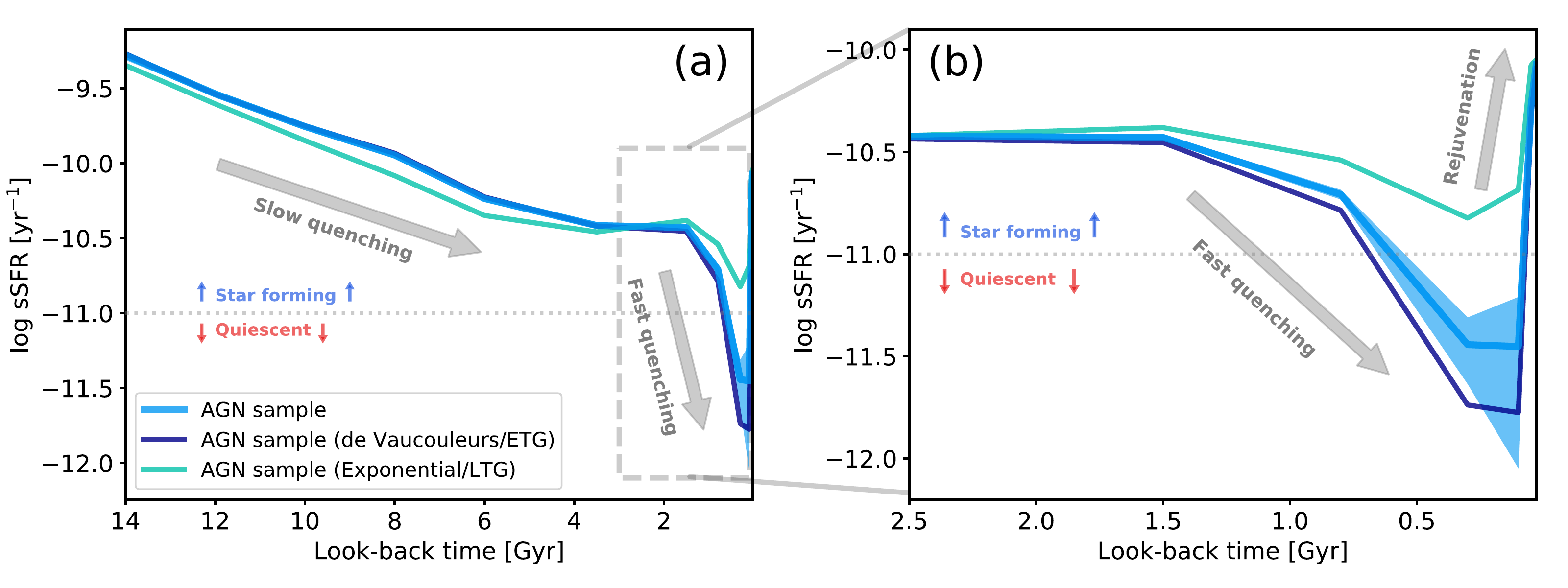}
    \end{center}
    \caption{{\bf Slow and rapid quenching in active galaxies.} Panel (a) shows how the initial and long-lasting quenching process in our AGN sample is followed by a fast quenching phase. Panel (b) is a zoom-in into the last 2.5 Gyr, revealing the morphological dependence of this fast quenching stage, which happens more abruptly in ETGs. The ubiquitous rejuvenation process becomes clear for ages younger than $\sim$ 0.1 Gyr. Shaded areas indicate the 1$\sigma$ confidence interval.}
    \label{fig:2}
\end{figure*}

\begin{figure*} 
    \begin{center}
    \includegraphics[width=8.5cm]{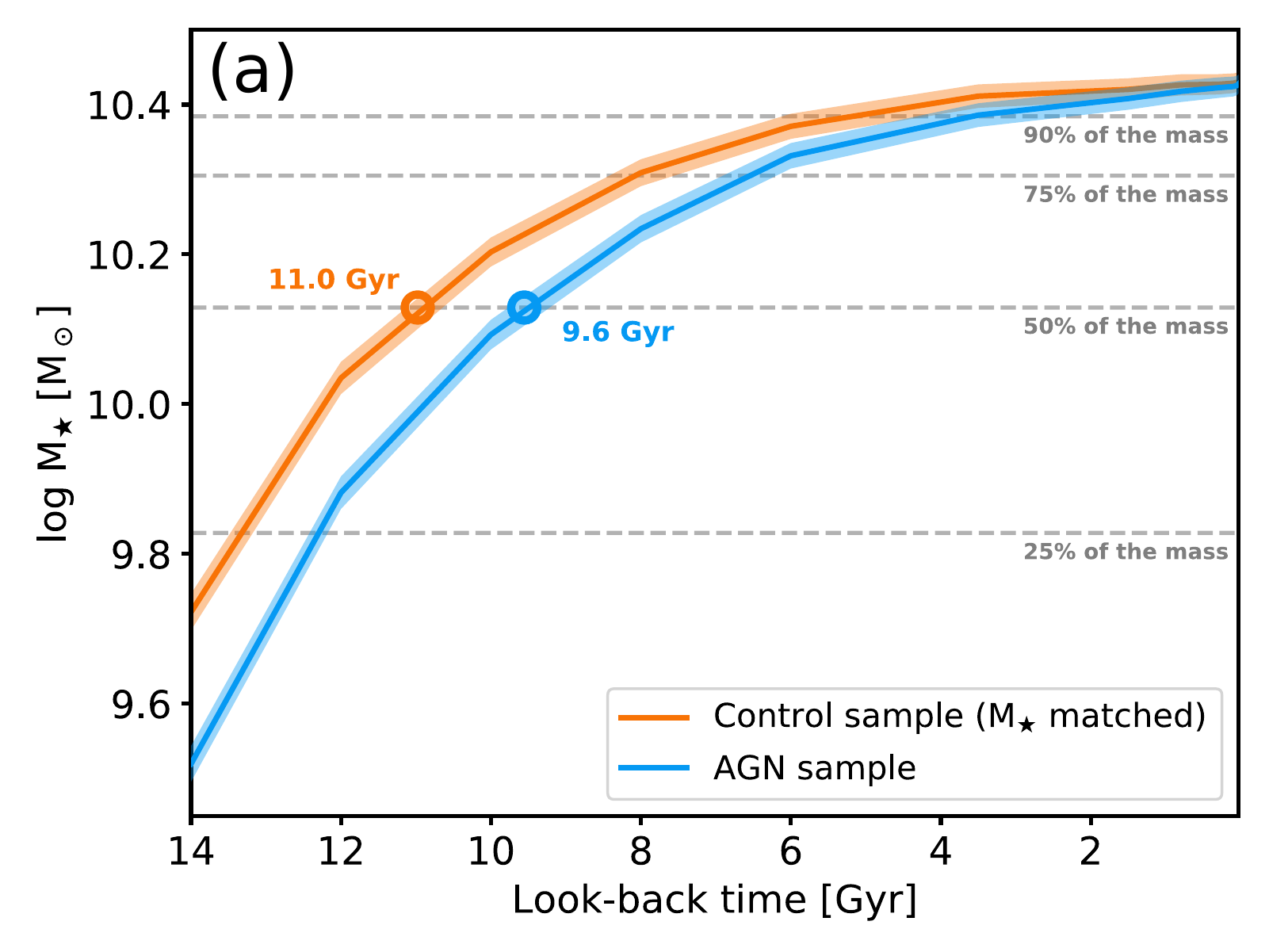}
    \includegraphics[width=8.5cm]{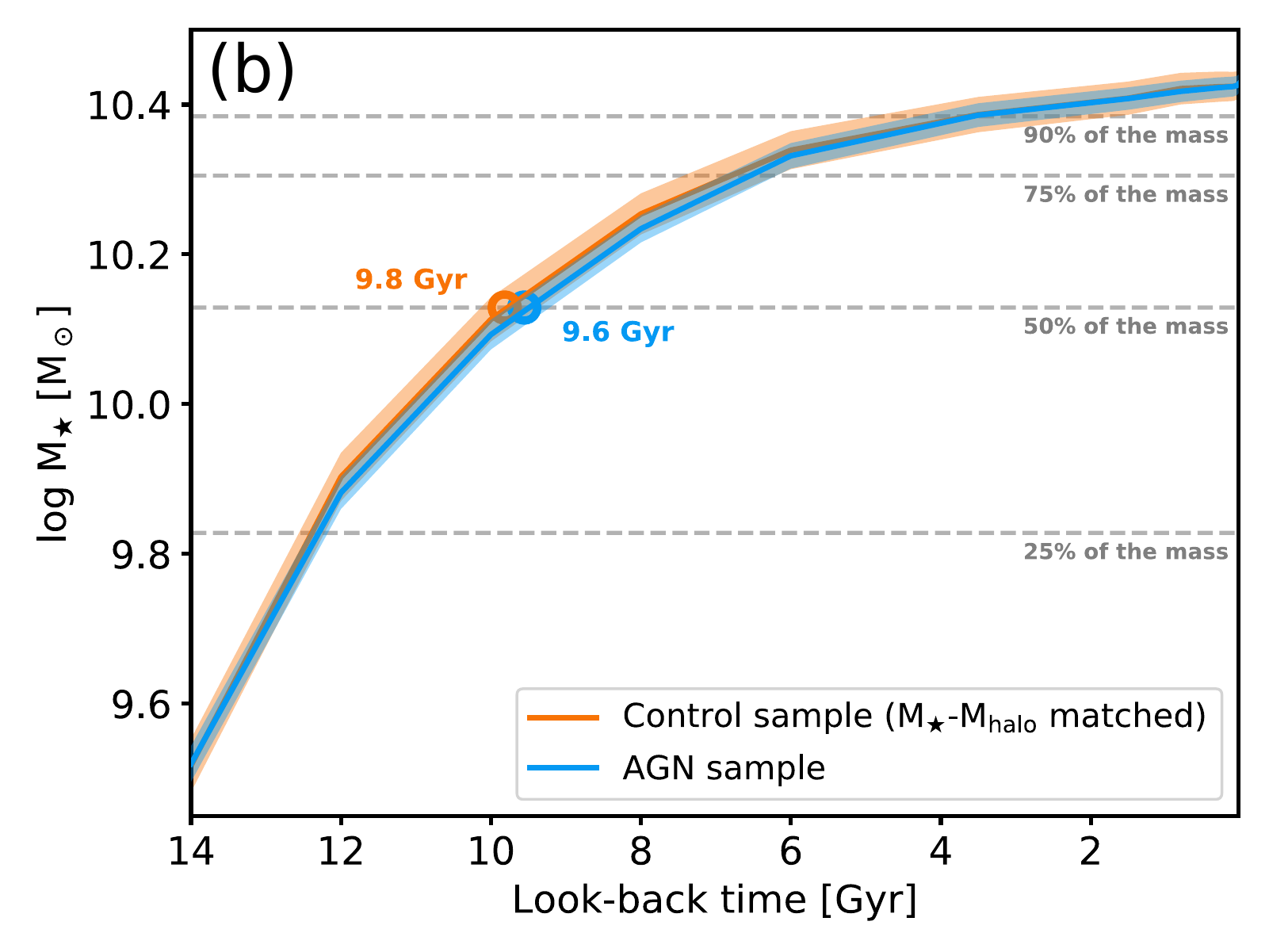}
    \end{center}
    \caption{{\bf Formation history comparison.} In Panel (a), the cumulative mass growth of our AGN sample (blue) is compared to our M$_\star$-matched control sample (orange). Due to the observed formation delay, AGN galaxies are younger than the average population with the same stellar mass. Similarly, panel (b) shows how the AGN sample compares to the control sample, this time matched in both stellar and halo masses. In this case, the formation history of AGN galaxies is indistinguishable from the control sample. Shaded areas indicate the 1$\sigma$ confidence interval and horizontal dashed lines mark when 25, 50, 75, and 90\% of the stellar mass was formed.}
    \label{fig:3}
\end{figure*}

Fig.~\ref{fig:1} shows the evolution of the specific star formation rate (sSFR) as a function of look-back time for our AGN sample (in blue) and our M$_\star$-matched control sample (in orange). Interestingly, the sSFR in both samples has been steadily decreasing for most of their (observed) evolution, reaching in both cases sSFR values consistent with a quiescent population ($\sim 10^{-11}$ yr$^{-1}$) until very recent epochs. However, a sudden increase in the sSFR at look-back times of $\sim$0.1 Gyr is observed in our sample of active galaxies, departing from the quiescent population at $z\sim0$.

While the logarithmic scale of Fig.~\ref{fig:1} facilitates the visualization of recent star formation episodes, quenching time-scales are more evident in linear units. In Fig.~\ref{fig:2} we show how the sSFR of the AGN population changes as a function of time, dividing our sample into later (LTG, in turquoise) and earlier (ETG, dark blue) morphological types, according to the SDSS imaging pipeline \citep{Stoughton02}. Two distinct phases in the evolution of the sSFR become clear from panel (a) in Fig.~\ref{fig:2}. An initial {\it slow quenching}, characterized by a steady decrease in the sSFR for around $\sim10$ Gyr ($\Delta$sSFR$/\Delta\mathrm{t}  \sim 0.1$ dex/Gyr), is followed by an abrupt {\it fast quenching} stage where the sSFR drops by up to an order of magnitude (i.e. $\Delta$sSFR$/\Delta\mathrm{t} \sim 1$ dex/Gyr). This {\it fast quenching} is responsible for driving the sSFR of our AGN sample from the star-forming to the quiescent population and it is mainly driven by the (dominant) ETG population. Panel (b) in Fig.~\ref{fig:2} reveals how {\it fast quenching} is morphology-dependent, as it happens faster in ETGs than in LTGs, in agreement with previous results \citep{Schawinski14}. For both ETGs and LTGs, the rejuvenation process showcased in Fig.~\ref{fig:1} is evident.

It is worth noting here that stellar evolution determines the time resolution of star formation histories measured from integrated spectra. Changes in the spectra of young stellar populations are much faster than in old stars. Hence, star formation events can only be measured when the time resolution of stellar population models is of the order of the star formation timescales, which effectively only happens at young ages. Additional rejuvenation events might have occurred in the past histories of these galaxies but they cannot be detected due to the coarse resolution of stellar population models at older ages. Note also that a change in the fitted wavelengths, in particular towards the blue, would only emphasize the observed rejuvenation as we would become more sensitive to younger stars. Moreover, it would not affect the comparison with the control sample as stellar masses and star formation histories are compared in relative terms. 

A close inspection reveals two key differences in the star formation histories of our AGN and control samples. First, the typical age of AGN-hosting galaxies is younger (9.6$\pm0.2$Gyr) than the average population of galaxies at the same mass range (11.0$\pm0.2$ Gyr). As shown in panel (a) of Fig.~\ref{fig:3}, this results from a delayed mass growth in AGN galaxies, unrelated to the rejuvenation episode which barely contributes to the actual mass growth. Second, the average halo mass of our AGN sample is also systematically higher than in the control sample ($\log$ M$_\mathrm{halo}$ = 12.45$\pm0.01$ and 12.30$\pm0.01$ \msun , respectively, where the confidence interval accounts for the statistical variance), similar to what it is found for radio-loud AGNs \citep{Mandelbaum09}. Albeit small, this offset in the average halo mass may have a direct effect on the formation histories of galaxies. Panel (b) of Fig.~\ref{fig:3} compares the formation history of our AGN sample with a control sample matched in both stellar and halo mass. The comparison between panels (a) and (b) in Fig.~\ref{fig:3} evidences a striking observational result: at fixed stellar mass, galaxies grew their stellar component at a different rate depending on the mass of their host halo. Furthermore, active galaxies are indistinguishable from the average population of galaxies with the same stellar and halo mass. 

\section{Discussion and Conclusions} \label{sec:discu}

\begin{figure} 
    \begin{center}
    \includegraphics[width=8.5cm]{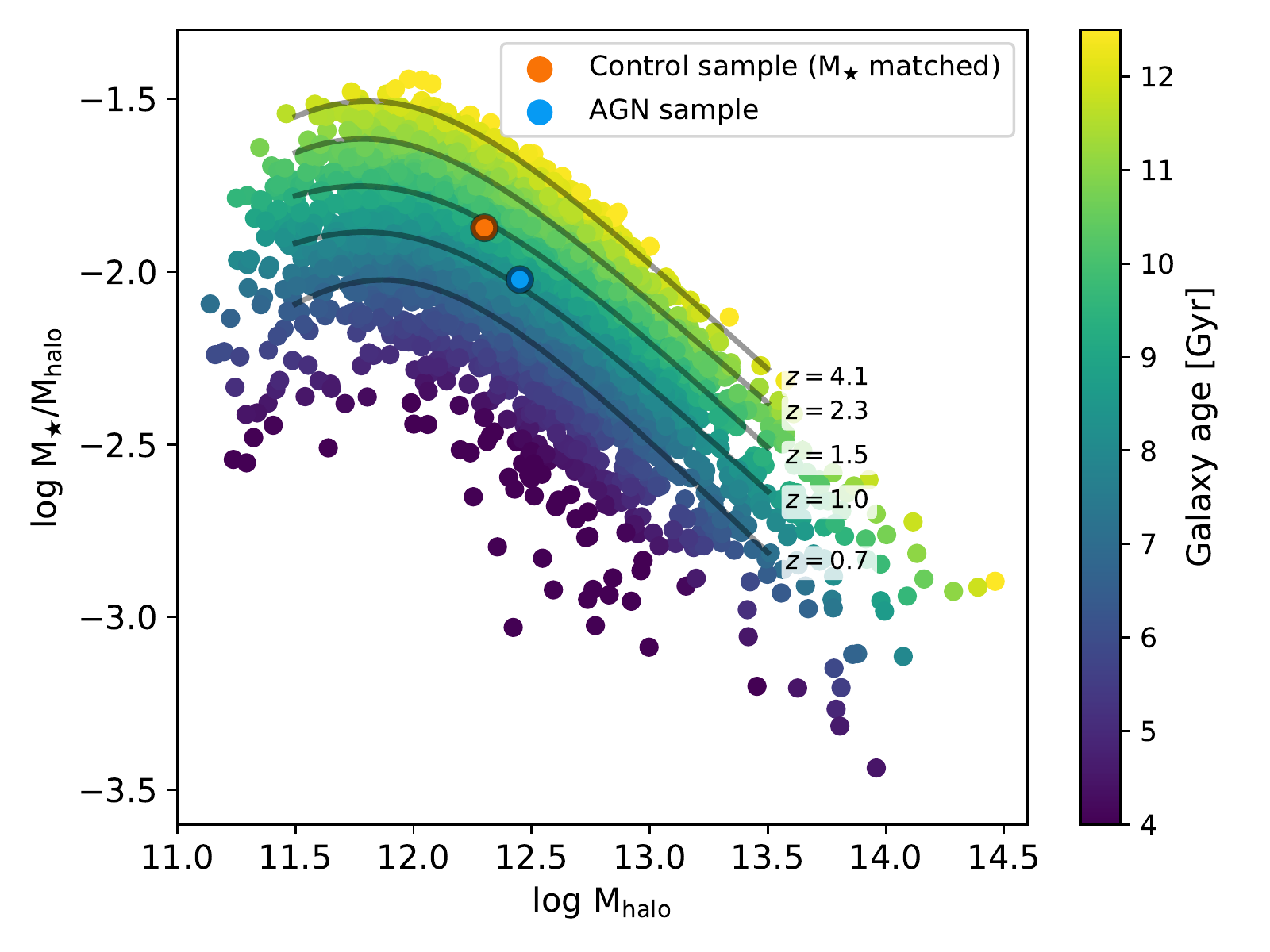}
    \end{center}
    \caption{{\bf Stellar-to-halo mass relation}. The SHMR is shown for our sample of AGN galaxies and the M$_\star$-matched control sample, color-coded by the age of the central galaxy. Lines of constant age are shown with a solid line, where ages have been translated into formation redshifts. Our sample of AGN galaxies (blue dot) is hosted by more massive halos than the M$_\star$-matched one (orange dot), which also results in AGN galaxies being younger than the average population at the same stellar mass.}
    \label{fig:4}
\end{figure}

The stellar-to-halo mass relation (SHMR) is a useful tool to probe galaxy evolution in a cosmological context, as it provides the average link between galaxies and host dark matter haloes which can then be tested against galaxy evolutionary models. Also the dispersion (scatter) around the mean of this relation has been suggested to correlate with the properties and formation history of the host dark matter halos \citep{Croton07}. Fig.~\ref{fig:4} shows the SHMR for our combined AGN and M$_\star$-matched control samples, color-coded by the age of the central galaxy. Interestingly, at fixed halo mass, galaxies with higher stellar masses tend to be older. Lines of constant age across the SHMR are marked with solid lines, where galaxy ages have been translated into formation redshifts.  The observed trend resembles the predicted correlation between halo formation time and the scatter in the SHMR, where galaxies with higher M$_\star$/M$_\mathrm{halo}$ are hosted by earlier-formed halos \citep{Matthee17}. Note that these differences in age across the SHMR also translate into a morphological separation, as ETGs tend to have older stellar population than LTGs.

Furthermore, Fig.~\ref{fig:4} helps to provide a coherent interpretation of the observed differences between our AGN and control samples. At fixed stellar mass, AGN galaxies are on average hosted by more massive halos which, as noted above, are thought to have formed later. Therefore, the delayed mass growth of AGN galaxies, and hence their younger ages, compared to the overall population of galaxies with the same stellar mass can be interpreted as the result of a late formation time of their host halos. In this scenario, AGN galaxies and their host halos would be systems less evolved than the mean population of galaxies in the same stellar mass range. Moreover, we propose that the rejuvenation process exhibited by AGN galaxies is also a consequence of their relatively younger evolutionary stage, as more gas is expected to be available than in more evolved halos. At fixed stellar and halo mass, local active and inactive galaxies show the same formation history, and whether a galaxy is optically classified as AGN or not likely relies on a stochastic gas infall process that is able to fuel the central black hole, leading as well to the observed rejuvenation phase \citep{McAlpine17}. Halo mass is therefore a critical quantity to understand the formation history of galaxies.

Our results have also direct implications on our understanding of the quenching process of massive galaxies. Given the average stellar and black hole mass of our sample (10.4 \msun  \ and 7.4 \msun, respectively), the expected AGN lifetime is $\sim10^8$ year \citep{Shankar04}, which is of the order of the rejuvenation time-scale as shown in Fig.~\ref{fig:1}. Therefore, our observations suggest that both AGN and star formation activity are triggered simultaneously. Positive black hole feedback might also contribute to the observed increase in the SFR but, given its frequency, geometry, and intensity \citep{Cresci15,Maiolino17}, it likely has a rather negligible impact in our measurements.

\begin{table*}
    \centering
    \begin{tabular}{cccccccc}
    \hline \hline
    ID & R.A. & Dec & M$_\bullet$ & M$_\mathrm{halo}$ & M$_\star$ & $\sigma_\mathrm{Reff}$  & Source \\
        & [Deg] & [Deg] & [$\log$ \msun] & [$\log$ \msun]  & [$\log$ \msun] & [\kms] & \\ 
    (1) & (2) & (3) & (4)  & (5)  & (6)  & (7) & (8) \\  
    \hline
    J152143.15+054033.9 & 230.429  &  5.676  & 9.18 & 12.34 & 10.64$\pm 0.03$  & 191.5 $\pm$ 4.6 & 5 \\
    ... & & & & & & & \\
    \hline
    \end{tabular}
\caption{ Main sample properties. (1) SDSS ID; (2), (3) Right ascension and declination (J2000); (4) Black hole mass, from the literature and corrected to the same virial factor; (5) Estimated halo mass, from the literature; (6) Stellar mass measurement based on the best-fitting stellar population properties derived from the optical SDSS spectrum; (7) Stellar velocity dispersion, measured from the SDSS optical spectrum and corrected to the effective radius of the galaxy; (8) Reference for the black hole mass measurement: \citep[0]{Reines15}, \citep[1]{Woo15}, \citep[2]{Greene07}, \citep[4]{Dong12}, \citep[5]{Liu19}, \citep[6]{Chilingarian18}.}   
\label{tab:1}

\end{table*}

Moreover, as revealed by Fig.~\ref{fig:2}, a prolonged star formation quenching preceded this onset of recent star formation and AGN activity in our sample. Note that this does not imply that AGN feedback is not responsible for quenching the star formation. Successive rejuvenation and nuclear activity episodes can be part of the natural quenching process of galaxies, heating up the gas within halos up until it becomes too hot to further sustain any steady star formation \citep{Bower17}. Such a delayed black hole-driven quenching is also supported by the observed coupled between nuclear outflows and star formation \citep{Cresci18}.

The same Fig.~\ref{fig:2} shows that, although an enhanced level of recent star formation is a common feature across our sample of broad-line AGN galaxies, it is more pronounced in ETGs, though it is also clearly present in LTGs. This sudden transition from a quiescent to a start-forming phase is usually referred to as {\it rejuvenation}. Rejuvenation may be interpreted as either a stochastic process in which both star formation and possibly AGN activity are triggered  \citep[e.g.][]{Mallmann18}, or a sudden increase in star formation specifically induced by AGN activity. We favour the former hypothesis as there is observational evidence for black hole fueling being indeed a rather stochastic process \citep[e.g.][]{Cristina17} and, in addition, there are also sporadic examples of LTGs within the matched galaxy sample with signs of rejuvenation in their recent star formation but no evident optical AGN signatures. Having uniform radio detections of local AGN would help in shedding light on the role of ``positive'' AGN feedback induced by AGN jets \citep[e.g.,][]{Gaibler19}. 

All in all, our results suggest that the popular interpretation of local AGN galaxies as a strictly transition phase between the star forming and quiescent population of galaxies may be incomplete  \citep[see e.g.][]{Sebastian18}, and that many AGN in the green valley could be actually moving towards (and not from) the main sequence. 

\section*{Data Availability}
Data access link http://skyserver.sdss.org/casjobs

\section*{Acknowledgements}
We would like to thank the referee for his/her useful feedback. IMN acknowledges support from grant PID2019-107427GB-C32 from the MCI. MM acknowledges support from the Beatriu de Pinos fellowship (2017-BP-00114) and from the Ramon y Cajal fellowship (RYC2019-027670-I). FS acknowledges partial support from a Leverhulme Trust Research Fellowship. For the purpose of open access, the authors have applied a CC-BY public copyright license to any author accepted manuscript version arising.




\bibliographystyle{mnras}
\bibliography{young} 

\begin{thebibliography}{}
\makeatletter
\relax
\def\mn@urlcharsother{\let\do\@makeother \do\$\do\&\do\#\do\^\do\_\do\%\do\~}
\def\mn@doi{\begingroup\mn@urlcharsother \@ifnextchar [ {\mn@doi@}
  {\mn@doi@[]}}
\def\mn@doi@[#1]#2{\def\@tempa{#1}\ifx\@tempa\@empty \href
  {http://dx.doi.org/#2} {doi:#2}\else \href {http://dx.doi.org/#2} {#1}\fi
  \endgroup}
\def\mn@eprint#1#2{\mn@eprint@#1:#2::\@nil}
\def\mn@eprint@arXiv#1{\href {http://arxiv.org/abs/#1} {{\tt arXiv:#1}}}
\def\mn@eprint@dblp#1{\href {http://dblp.uni-trier.de/rec/bibtex/#1.xml}
  {dblp:#1}}
\def\mn@eprint@#1:#2:#3:#4\@nil{\def\@tempa {#1}\def\@tempb {#2}\def\@tempc
  {#3}\ifx \@tempc \@empty \let \@tempc \@tempb \let \@tempb \@tempa \fi \ifx
  \@tempb \@empty \def\@tempb {arXiv}\fi \@ifundefined
  {mn@eprint@\@tempb}{\@tempb:\@tempc}{\expandafter \expandafter \csname
  mn@eprint@\@tempb\endcsname \expandafter{\@tempc}}}

\bibitem[\protect\citeauthoryear{{Ahn} et~al.,}{{Ahn} et~al.}{2014}]{SDSS10}
{Ahn} C.~P.,  et~al., 2014, \mn@doi [\apjs] {10.1088/0067-0049/211/2/17}, \href
  {https://ui.adsabs.harvard.edu/abs/2014ApJS..211...17A} {211, 17}

\bibitem[\protect\citeauthoryear{{Angthopo}, {Ferreras}  \& {Silk}}{{Angthopo}
  et~al.}{2020}]{Angthopo20}
{Angthopo} J.,  {Ferreras} I.,   {Silk} J.,  2020, \mn@doi [\mnras]
  {10.1093/mnras/staa1276}, \href
  {https://ui.adsabs.harvard.edu/abs/2020MNRAS.495.2720A} {495, 2720}

\bibitem[\protect\citeauthoryear{{Blanton} et~al.,}{{Blanton}
  et~al.}{2005}]{Blanton05}
{Blanton} M.~R.,  et~al., 2005, \mn@doi [\aj] {10.1086/429803}, \href
  {https://ui.adsabs.harvard.edu/abs/2005AJ....129.2562B} {129, 2562}

\bibitem[\protect\citeauthoryear{{Bluck} et~al.,}{{Bluck}
  et~al.}{2020}]{Bluck20}
{Bluck} A. F.~L.,  et~al., 2020, \mn@doi [\mnras] {10.1093/mnras/staa2806},
  \href {https://ui.adsabs.harvard.edu/abs/2020MNRAS.499..230B} {499, 230}

\bibitem[\protect\citeauthoryear{{Bower}, {Schaye}, {Frenk}, {Theuns},
  {Schaller}, {Crain}  \& {McAlpine}}{{Bower} et~al.}{2017}]{Bower17}
{Bower} R.~G.,  {Schaye} J.,  {Frenk} C.~S.,  {Theuns} T.,  {Schaller} M.,
  {Crain} R.~A.,   {McAlpine} S.,  2017, \mn@doi [\mnras]
  {10.1093/mnras/stw2735}, \href
  {https://ui.adsabs.harvard.edu/abs/2017MNRAS.465...32B} {465, 32}

\bibitem[\protect\citeauthoryear{{Cappellari} \& {Emsellem}}{{Cappellari} \&
  {Emsellem}}{2004}]{ppxf}
{Cappellari} M.,  {Emsellem} E.,  2004, \mn@doi [\pasp] {10.1086/381875}, \href
  {https://ui.adsabs.harvard.edu/abs/2004PASP..116..138C} {116, 138}

\bibitem[\protect\citeauthoryear{{Chilingarian}, {Katkov}, {Zolotukhin},
  {Grishin}, {Beletsky}, {Boutsia}  \& {Osip}}{{Chilingarian}
  et~al.}{2018}]{Chilingarian18}
{Chilingarian} I.~V.,  {Katkov} I.~Y.,  {Zolotukhin} I.~Y.,  {Grishin} K.~A.,
  {Beletsky} Y.,  {Boutsia} K.,   {Osip} D.~J.,  2018, \mn@doi [\apj]
  {10.3847/1538-4357/aad184}, \href
  {https://ui.adsabs.harvard.edu/abs/2018ApJ...863....1C} {863, 1}

\bibitem[\protect\citeauthoryear{{Conroy} \& {van Dokkum}}{{Conroy} \& {van
  Dokkum}}{2012}]{conroy}
{Conroy} C.,  {van Dokkum} P.,  2012, \mn@doi [\apj]
  {10.1088/0004-637X/747/1/69}, \href
  {http://adsabs.harvard.edu/abs/2012ApJ...747...69C} {747, 69}

\bibitem[\protect\citeauthoryear{{Cresci} \& {Maiolino}}{{Cresci} \&
  {Maiolino}}{2018}]{Cresci18}
{Cresci} G.,  {Maiolino} R.,  2018, \mn@doi [Nature Astronomy]
  {10.1038/s41550-018-0404-5}, \href
  {https://ui.adsabs.harvard.edu/abs/2018NatAs...2..179C} {2, 179}

\bibitem[\protect\citeauthoryear{{Cresci} et~al.,}{{Cresci}
  et~al.}{2015}]{Cresci15}
{Cresci} G.,  et~al., 2015, \mn@doi [\aap] {10.1051/0004-6361/201526581}, \href
  {https://ui.adsabs.harvard.edu/abs/2015A&A...582A..63C} {582, A63}

\bibitem[\protect\citeauthoryear{{Croton}, {Gao}  \& {White}}{{Croton}
  et~al.}{2007}]{Croton07}
{Croton} D.~J.,  {Gao} L.,   {White} S. D.~M.,  2007, \mn@doi [\mnras]
  {10.1111/j.1365-2966.2006.11230.x}, \href
  {https://ui.adsabs.harvard.edu/abs/2007MNRAS.374.1303C} {374, 1303}

\bibitem[\protect\citeauthoryear{Davé, Crain, Stevens, Narayanan, Saintonge,
  Catinella  \& Cortese}{Davé et~al.}{2020}]{Dav2020}
Davé R.,  Crain R.~A.,  Stevens A. R.~H.,  Narayanan D.,  Saintonge A.,
  Catinella B.,   Cortese L.,  2020, \mn@doi [Monthly Notices of the Royal
  Astronomical Society] {10.1093/mnras/staa1894}, 497, 146–166

\bibitem[\protect\citeauthoryear{{Dong}, {Ho}, {Yuan}, {Wang}, {Fan}, {Zhou}
  \& {Jiang}}{{Dong} et~al.}{2012}]{Dong12}
{Dong} X.-B.,  {Ho} L.~C.,  {Yuan} W.,  {Wang} T.-G.,  {Fan} X.,  {Zhou} H.,
  {Jiang} N.,  2012, \mn@doi [\apj] {10.1088/0004-637X/755/2/167}, \href
  {https://ui.adsabs.harvard.edu/abs/2012ApJ...755..167D} {755, 167}

\bibitem[\protect\citeauthoryear{{Gaibler}, {Khochfar}, {Krause}  \&
  {Silk}}{{Gaibler} et~al.}{2012}]{Gaibler19}
{Gaibler} V.,  {Khochfar} S.,  {Krause} M.,   {Silk} J.,  2012, \mn@doi
  [\mnras] {10.1111/j.1365-2966.2012.21479.x}, \href
  {https://ui.adsabs.harvard.edu/abs/2012MNRAS.425..438G} {425, 438}

\bibitem[\protect\citeauthoryear{{Greene} \& {Ho}}{{Greene} \&
  {Ho}}{2007}]{Greene07}
{Greene} J.~E.,  {Ho} L.~C.,  2007, \mn@doi [\apj] {10.1086/522082}, \href
  {https://ui.adsabs.harvard.edu/abs/2007ApJ...670...92G} {670, 92}

\bibitem[\protect\citeauthoryear{{Kova{\v{c}}evi{\'c}}, {Popovi{\'c}}  \&
  {Dimitrijevi{\'c}}}{{Kova{\v{c}}evi{\'c}} et~al.}{2010}]{feii}
{Kova{\v{c}}evi{\'c}} J.,  {Popovi{\'c}} L.~{\v{C}}.,   {Dimitrijevi{\'c}}
  M.~S.,  2010, \mn@doi [\apjs] {10.1088/0067-0049/189/1/15}, \href
  {https://ui.adsabs.harvard.edu/abs/2010ApJS..189...15K} {189, 15}

\bibitem[\protect\citeauthoryear{{Lim}, {Mo}, {Lu}, {Wang}  \& {Yang}}{{Lim}
  et~al.}{2017}]{Lim17}
{Lim} S.~H.,  {Mo} H.~J.,  {Lu} Y.,  {Wang} H.,   {Yang} X.,  2017, \mn@doi
  [\mnras] {10.1093/mnras/stx1462}, \href
  {https://ui.adsabs.harvard.edu/abs/2017MNRAS.470.2982L} {470, 2982}

\bibitem[\protect\citeauthoryear{{Liu}, {Liu}, {Dong}, {Zhou}, {Wang}, {Lu}  \&
  {Yuan}}{{Liu} et~al.}{2019}]{Liu19}
{Liu} H.-Y.,  {Liu} W.-J.,  {Dong} X.-B.,  {Zhou} H.,  {Wang} T.,  {Lu} H.,
  {Yuan} W.,  2019, \mn@doi [\apjs] {10.3847/1538-4365/ab298b}, \href
  {https://ui.adsabs.harvard.edu/abs/2019ApJS..243...21L} {243, 21}

\bibitem[\protect\citeauthoryear{{Maiolino} et~al.,}{{Maiolino}
  et~al.}{2017}]{Maiolino17}
{Maiolino} R.,  et~al., 2017, \mn@doi [\nat] {10.1038/nature21677}, \href
  {https://ui.adsabs.harvard.edu/abs/2017Natur.544..202M} {544, 202}

\bibitem[\protect\citeauthoryear{{Mallmann} et~al.,}{{Mallmann}
  et~al.}{2018}]{Mallmann18}
{Mallmann} N.~D.,  et~al., 2018, \mn@doi [\mnras] {10.1093/mnras/sty1364},
  \href {https://ui.adsabs.harvard.edu/abs/2018MNRAS.478.5491M} {478, 5491}

\bibitem[\protect\citeauthoryear{{Mandelbaum}, {Li}, {Kauffmann}  \&
  {White}}{{Mandelbaum} et~al.}{2009}]{Mandelbaum09}
{Mandelbaum} R.,  {Li} C.,  {Kauffmann} G.,   {White} S. D.~M.,  2009, \mn@doi
  [\mnras] {10.1111/j.1365-2966.2008.14235.x}, \href
  {https://ui.adsabs.harvard.edu/abs/2009MNRAS.393..377M} {393, 377}

\bibitem[\protect\citeauthoryear{{Maraston}}{{Maraston}}{2005}]{Maraston05}
{Maraston} C.,  2005, \mn@doi [\mnras] {10.1111/j.1365-2966.2005.09270.x},
  \href {https://ui.adsabs.harvard.edu/abs/2005MNRAS.362..799M} {362, 799}

\bibitem[\protect\citeauthoryear{{Matthee}, {Schaye}, {Crain}, {Schaller},
  {Bower}  \& {Theuns}}{{Matthee} et~al.}{2017}]{Matthee17}
{Matthee} J.,  {Schaye} J.,  {Crain} R.~A.,  {Schaller} M.,  {Bower} R.,
  {Theuns} T.,  2017, \mn@doi [\mnras] {10.1093/mnras/stw2884}, \href
  {https://ui.adsabs.harvard.edu/abs/2017MNRAS.465.2381M} {465, 2381}

\bibitem[\protect\citeauthoryear{{McAlpine}, {Bower}, {Harrison}, {Crain},
  {Schaller}, {Schaye}  \& {Theuns}}{{McAlpine} et~al.}{2017}]{McAlpine17}
{McAlpine} S.,  {Bower} R.~G.,  {Harrison} C.~M.,  {Crain} R.~A.,  {Schaller}
  M.,  {Schaye} J.,   {Theuns} T.,  2017, \mn@doi [\mnras]
  {10.1093/mnras/stx658}, \href
  {https://ui.adsabs.harvard.edu/abs/2017MNRAS.468.3395M} {468, 3395}

\bibitem[\protect\citeauthoryear{{Mitchell}, {Schaye}  \& {Bower}}{{Mitchell}
  et~al.}{2020}]{Mitchell20}
{Mitchell} P.~D.,  {Schaye} J.,   {Bower} R.~G.,  2020, \mn@doi [\mnras]
  {10.1093/mnras/staa2252}, \href
  {https://ui.adsabs.harvard.edu/abs/2020MNRAS.497.4495M} {497, 4495}

\bibitem[\protect\citeauthoryear{{Nelson} et~al.,}{{Nelson}
  et~al.}{2019}]{Nelson19}
{Nelson} D.,  et~al., 2019, \mn@doi [\mnras] {10.1093/mnras/stz2306}, \href
  {https://ui.adsabs.harvard.edu/abs/2019MNRAS.490.3234N} {490, 3234}

\bibitem[\protect\citeauthoryear{{Padmanabhan} et~al.,}{{Padmanabhan}
  et~al.}{2008}]{Padmanabhan08}
{Padmanabhan} N.,  et~al., 2008, \mn@doi [\apj] {10.1086/524677}, \href
  {https://ui.adsabs.harvard.edu/abs/2008ApJ...674.1217P} {674, 1217}

\bibitem[\protect\citeauthoryear{{Ramos Almeida} \& {Ricci}}{{Ramos Almeida} \&
  {Ricci}}{2017}]{Cristina17}
{Ramos Almeida} C.,  {Ricci} C.,  2017, \mn@doi [Nature Astronomy]
  {10.1038/s41550-017-0232-z}, \href
  {https://ui.adsabs.harvard.edu/abs/2017NatAs...1..679R} {1, 679}

\bibitem[\protect\citeauthoryear{{Reines} \& {Volonteri}}{{Reines} \&
  {Volonteri}}{2015}]{Reines15}
{Reines} A.~E.,  {Volonteri} M.,  2015, \mn@doi [\apj]
  {10.1088/0004-637X/813/2/82}, \href
  {https://ui.adsabs.harvard.edu/abs/2015ApJ...813...82R} {813, 82}

\bibitem[\protect\citeauthoryear{{S{\'a}nchez} et~al.,}{{S{\'a}nchez}
  et~al.}{2018}]{Sebastian18}
{S{\'a}nchez} S.~F.,  et~al., 2018, \rmxaa, \href
  {https://ui.adsabs.harvard.edu/abs/2018RMxAA..54..217S} {54, 217}

\bibitem[\protect\citeauthoryear{{Schawinski}, {Thomas}, {Sarzi}, {Maraston},
  {Kaviraj}, {Joo}, {Yi}  \& {Silk}}{{Schawinski} et~al.}{2007}]{Schawinski07}
{Schawinski} K.,  {Thomas} D.,  {Sarzi} M.,  {Maraston} C.,  {Kaviraj} S.,
  {Joo} S.-J.,  {Yi} S.~K.,   {Silk} J.,  2007, \mn@doi [\mnras]
  {10.1111/j.1365-2966.2007.12487.x}, \href
  {https://ui.adsabs.harvard.edu/abs/2007MNRAS.382.1415S} {382, 1415}

\bibitem[\protect\citeauthoryear{{Schawinski} et~al.,}{{Schawinski}
  et~al.}{2014}]{Schawinski14}
{Schawinski} K.,  et~al., 2014, \mn@doi [\mnras] {10.1093/mnras/stu327}, \href
  {https://ui.adsabs.harvard.edu/abs/2014MNRAS.440..889S} {440, 889}

\bibitem[\protect\citeauthoryear{{Shankar}, {Salucci}, {Granato}, {De Zotti}
  \& {Danese}}{{Shankar} et~al.}{2004}]{Shankar04}
{Shankar} F.,  {Salucci} P.,  {Granato} G.~L.,  {De Zotti} G.,   {Danese} L.,
  2004, \mn@doi [\mnras] {10.1111/j.1365-2966.2004.08261.x}, \href
  {https://ui.adsabs.harvard.edu/abs/2004MNRAS.354.1020S} {354, 1020}

\bibitem[\protect\citeauthoryear{{Stoughton} et~al.,}{{Stoughton}
  et~al.}{2002}]{Stoughton02}
{Stoughton} C.,  et~al., 2002, \mn@doi [\aj] {10.1086/324741}, \href
  {https://ui.adsabs.harvard.edu/abs/2002AJ....123..485S} {123, 485}

\bibitem[\protect\citeauthoryear{{Terrazas} et~al.,}{{Terrazas}
  et~al.}{2020}]{Terrazas20}
{Terrazas} B.~A.,  et~al., 2020, \mn@doi [\mnras] {10.1093/mnras/staa374},
  \href {https://ui.adsabs.harvard.edu/abs/2020MNRAS.493.1888T} {493, 1888}

\bibitem[\protect\citeauthoryear{{Vazdekis} et~al.,}{{Vazdekis}
  et~al.}{2015}]{Vazdekis15}
{Vazdekis} A.,  et~al., 2015, \mn@doi [\mnras] {10.1093/mnras/stv151}, \href
  {https://ui.adsabs.harvard.edu/abs/2015MNRAS.449.1177V} {449, 1177}

\bibitem[\protect\citeauthoryear{{Woo}, {Yoon}, {Park}, {Park}  \& {Kim}}{{Woo}
  et~al.}{2015}]{Woo15}
{Woo} J.-H.,  {Yoon} Y.,  {Park} S.,  {Park} D.,   {Kim} S.~C.,  2015, \mn@doi
  [\apj] {10.1088/0004-637X/801/1/38}, \href
  {https://ui.adsabs.harvard.edu/abs/2015ApJ...801...38W} {801, 38}

\makeatother
\end{thebibliography}


\bsp	
\label{lastpage}
\end{document}